\documentclass[aps,pre,superscriptaddress,reprint]{revtex4-1}
\usepackage[dvips]{graphicx}
\usepackage{amssymb,amsfonts,amsmath}
\usepackage{color}
\usepackage{ulem}
\usepackage[hidelinks]{hyperref}

\begin{document}
\title{Hard topological versus soft geometrical magnetic
particle transport}

\author{Anna M. E. B. Rossi}
\affiliation{Experimentalphysik V, Physikalisches Institut,
 Universit{\"a}t Bayreuth, D-95440 Bayreuth, Germany}

\author{Jonas Bugase}
\affiliation{Experimentalphysik V, Physikalisches Institut,
 Universit{\"a}t Bayreuth, D-95440 Bayreuth, Germany}

\author{Thomas Lachner}
\affiliation{Experimentalphysik V, Physikalisches Institut,
 Universit{\"a}t Bayreuth, D-95440 Bayreuth, Germany}

\author{Adrian Ernst}
\affiliation{Experimentalphysik V, Physikalisches Institut,
 Universit{\"a}t Bayreuth, D-95440 Bayreuth, Germany}

\author{Daniel de las Heras}
\affiliation{Theoretische Physik II, Physikalisches Institut,
 Universit{\"a}t Bayreuth, D-95440 Bayreuth, Germany}

\author{Thomas M. Fischer}
\email{thomas.fischer@uni-bayreuth.de}
\affiliation{Experimentalphysik V, Physikalisches Institut,
 Universit{\"a}t Bayreuth, D-95440 Bayreuth, Germany}

\date{\today}
 
\begin{abstract}
The transport on top of a periodic two-dimensional hexagonal magnetic pattern
of (i) a single macroscopic steel sphere, (ii) a doublet of wax/magnetite
composite spheres, and (iii) an immiscible mixture of ferrofluid droplets with a
perfluorinated liquid is studied experimentally and analyzed theoretically.
The transport of all these magnetic objects is achieved by moving an external
permanent magnet on a closed modulation loop around the two-dimensional magnetic pattern.
The transport of one and also that of two objects per unit cell is topologically protected
and characterized by discrete displacements of the particles as we continuously scan
through a family of modulation loops. The direction and the type of transport
is characterized by the winding numbers of the modulation loops around
special objects in control space, which is the space of possible directions
of the external magnetic field.
The number of winding numbers necessary for characterizing the topological transport
increases with the number of particles per unit cell.
The topological character of the transport is  destroyed when transporting a large collection
of particles per unit cell like it is the case for a macroscopic assembly of magnetic nano particles
in a ferrofluid droplet for which the transport is geometrical and no longer topological. 
To characterize the change in the transport from topological to geometrical, we perform computer
simulations of the transport of an increasing number of particles per unit cell.
\end{abstract}

\maketitle

\section{Introduction}\label{Introduction}
When a system is driven adiabatically, its state changes slower than any
relaxation time. A state of a classical system then follows the same path
independently of the speed of the driving. If driven adiabatically at different
speeds, a state of a quantum system also
follows the same path up to a global dynamic phase~\cite{TI} of its wave 
function that cannot be measured. Measurable quantities are geometrical
in the adiabatic limit since they can be deduced from the path without knowledge
of the particular time table with which one drives the system along this
path. For a periodically driven system the transport of particles over a
period then is proportional to a geometric quantity of the loop of the
driving field~\cite{GPP1,Berry,Hannay}. For example, autonomous low Reynolds number
swimmers propel by a distance proportional to the area of the driving control loop
in shape space~\cite{Wilczek1,Wilczek2}. Adiabatic quantum two-band electrons
propel by an amount proportional to the area enclosed by the SU(2) respectively
SO(3)$\cong $SU(2)$/\mathbb{Z}_2$ matrices of the periodic control loop induced by
the external field~\cite{Hasan,TI}.
The area of a loop is a geometric quantity that continuously changes when
the driving control loop is altered.  When symmetries or other conditions constraint
the driving loop, geometrical properties might become discrete global properties
called topological invariants. Then, the transport no longer changes continuously
with the loop since families of loops share the same topological invariant. 
The transport changes discretely between two families of 
loops with different values of the topological invariant~\cite{Hasan}. The transport is robust because it is topologically
protected. For example, in a nucleus one nucleon must rotate by multiples of $2\pi$ when it
propels by one lattice constant above the lattice of the crystallized nucleons
that form the rest of the nucleus~\cite{Harland2018}.    

Understanding how a system changes from geometrical towards topological
is important because it might help to change a geometrical and thus
fragile system towards a topological and thus robust system. For example, in the quantum
Hall effect steps between the plateaus in the conductivity can be created
either by lowering the temperature or by using clean systems with fewer
impurities~\cite{Klitzing}. Both methods decrease the probability of exciting
unoccupied bulk Landau levels and thus make the system topological. For many
quantum and classical systems~\cite{Mao,Paulose,Rechtsman,Nash,Kane,Huber}
the transition from geometric towards topological transport can
be understood via the amount of dissipation occurring due to the scattering
between states. It has, however, been shown that there exist non-Hermitian
quantum and dissipative classical topological transport
systems~\cite{Murugan,Xiong,Yao,Kunst,Loehr} where it is precisely the dissipation that causes
the topological character of the transport. For these systems the transition
from topological towards geometrical must be different.

Here, we show experimentally and with computer simulations
that a macroscopic topological magnetic particle pump~\cite{Rossi}, that
transports paramagnetic or soft magnetic particles across a magnetic
lattice, is topological when transporting a small number of particles per unit
cell. The transport is robust for those modulation loops
of a driving homogeneous external field that share the same topological
invariant. 
Subclasses of modulation loops appear for a loading with two or more particles per
unit cell increasing the number of discrete steps.
However, for loadings with a macroscopic ensemble of magnetic nanoparticles,
such as a ferrofluid droplet, the topological nature of the transport is
destroyed and becomes geometrical.

\section{Topogeometrical pump}
\begin{figure*}
	\includegraphics[width=2\columnwidth]{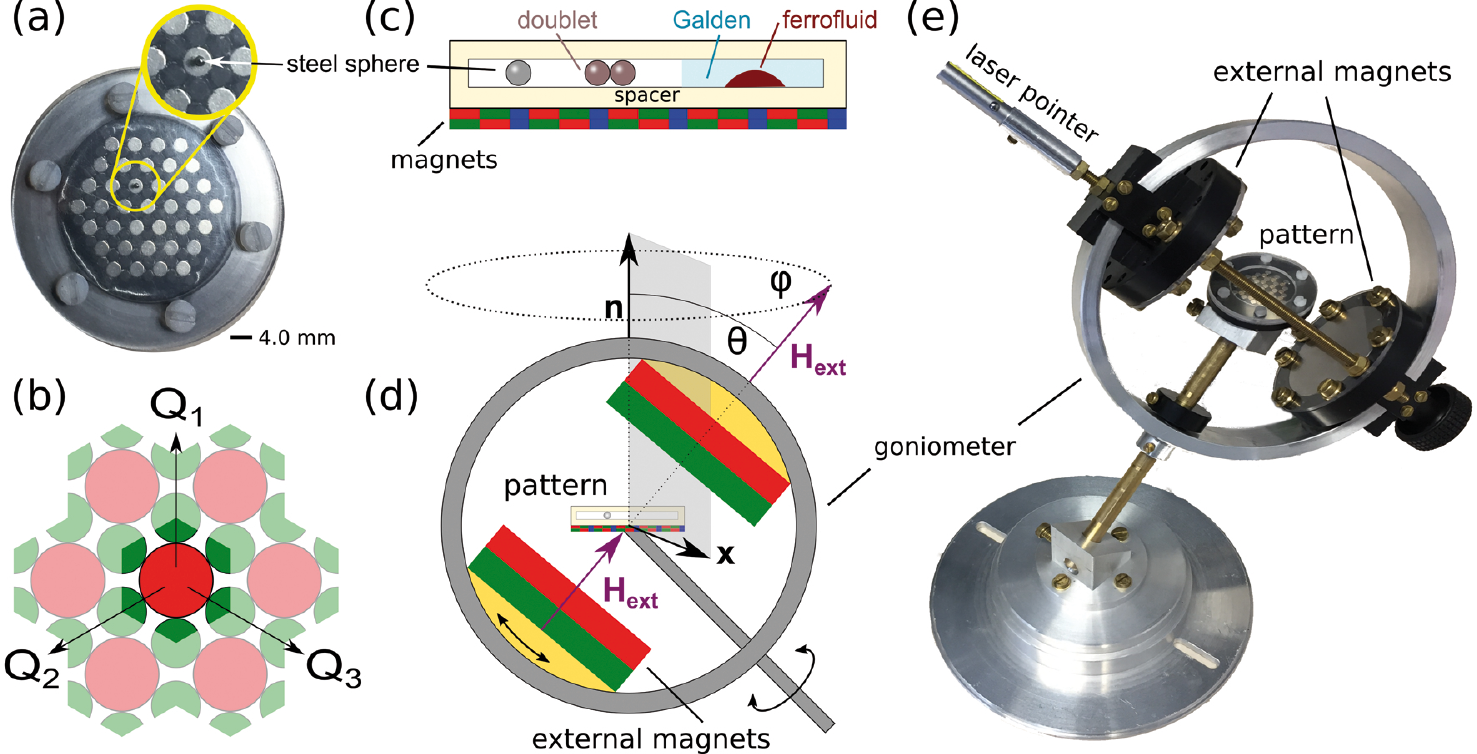}
	\caption
	{(a) Top view of the hexagonal magnetic pattern. The inset
		is a close view of the transported steel sphere.  (b) Scheme of the
		position and the orientation of the magnets.
		Silver areas in the sample (red areas
		in the scheme) are magnetized up. Black (green) areas are
		magnetized down. One unit cell is emphasized in full
		colors. The vector $\mathbf{Q}_1$ is one of the primitive reciprocal
		lattice vectors. (c) Side view of the pattern and the compartment
		holding either one steel sphere, two wax spheres, or Galden and ferrofluid.
		(d) Schematic of the goniometer and the external magnets surrounding the sample. (e) A
		photo of the setup.}
	\label{fig1}
\end{figure*}

The system consists of a two-dimensional hexagonal magnetic pattern made
of up- and down-magnetized magnets, see Fig.~\ref{fig1}a.
The pattern creates a two-dimensional
magnetic potential that acts on paramagnetic objects located above the pattern
at fixed elevation. The potential is a function of the position
$\mathbf x_{\cal A}\in\cal A$ of the paramagnetic object in action space $\cal A$
, which is the plane parallel to the pattern in which the objects are located.
A uniform external magnetic field is also applied to the system. Hence,
the total potential depends parametrically on the direction of the superimposed
external magnetic field. Paramagnetic objects, such as soft magnetic
spheres and ferrofluid droplets move in action space when we
adiabatically modulate the total potential by changing the direction of the
uniform external field. 

Our two dimensional magnetic hexagonal lattice is built from an arrangement
of NbB-magnets~\cite{Rossi}.  The lattice (Fig.~\ref{fig1}a) consists of
large (l) and small (s) cylindrical magnets of height $h=2$ mm,
diameters $d_l=3$~mm and $d_s=2$~mm, and remanences
$\mu_0 M_l= 1.19$~T  and $\mu_0 M_s= 1.35$~T, with $\mu_0$ 
the permeability of free space. The large
magnets are magnetized upwards (silver magnets in Fig~\ref{fig1}a and red regions
in Fig.~\ref{fig1}b) and are surrounded by six small magnets that are
magnetized downwards (black magnets in Fig~\ref{fig1}a and green regions in Fig.~\ref{fig1}b)
and that touch the large magnets. The
primitive unit cell of the lattice is a six fold symmetric $C_6$ hexagon
with corners centered within the smaller magnets, see Fig.~\ref{fig1}b.
Each unit cell thus contains one large magnet and two small magnets. 
The resulting lattice with two primitive lattice vectors of length $a=4.33$~mm
is mechanically metastable
in zero external
magnetic field. Therefore, we need to fix the metastable arrangement with
an epoxy resin placed in the voids and in the two dimensional surroundings
of the pattern. This stabilizes the pattern also in the presence of an external field.
The pattern is put on a support and covered with a transparent PMMA and
white-illuminated spacer of thickness $z=1.5$~mm  (Fig.~\ref{fig1}c).

The total magnetic field is the sum of the pattern ${\mathbf H}_p$ and the external
${\mathbf H}_{\text ext}$ contributions
\begin{equation}
{\mathbf H}={\mathbf H}_p+{\mathbf H}_\text{ext}.
\end{equation}
The potential energy of a paramagnetic object in the total magnetic field ${\rm H}$
is proportional to the square of the magnetic field 
\begin{equation}
U(\mathbf x_{\cal A})\propto - {\mathbf H}^2,\label{EQpotential}
\end{equation}
and it can be decomposed into a discrete Fourier series of contributions from
reciprocal lattice vectors~\cite{Loehr,delasHeras}. 
The Fourier series of the potential evaluated in a plane above the pattern and parallel to it
is the square of the Fourier series of the magnetization of the
pattern augmented by the external field. As a function of the elevation, the
higher Fourier coefficients are attenuated more than those with
lower reciprocal vectors. At the experimental elevation (comparable to the length
of the unit cell of the pattern), only the "universal"
contributions to the potential from the lowest non-zero reciprocal lattice
vectors remain relevant~\cite{Loehr,delasHeras}. The purpose of the spacer (Fig.~\ref{fig1}c)
is thus to render the potential universal such that only the symmetry and
not the fine details of the pattern are important.

We either place one steel sphere of diameter $2r=1$~mm or two spheres
of diameter $2r=0.5$~mm consisting of $10:1$ weight percent
mixture of wax and magnetite on top of the spacer. Alternatively, we spray the spacer with PTFE, place two fluids, a nonmagnetic fluid (Galden), and an aqueous ferrofluid
immiscible with the Galden at a volume ratio Galden/ferrofluid of $152:1$ on top of the PTFE and close the compartment with a transparent lid.
The magnetic pattern with the transported paramagnetic object on top is then placed
in the center of a goniometer which is set up at an angle of $45$ degrees to ensure
that the relevant motion is not affected by the restrictions of motion of the goniometer 
caused by the support. Both a sketch and an actual picture of the setup are shown in
Figs.~\ref{fig1}d and~\ref{fig1}e, respectively.
The goniometer holds two NbB-magnets
(diameter $d_\text{ext}=60$~mm, thickness  $t_\text{ext}=10$~mm, and remanence $\mu_0 M_\text{ext}=1.28$~T)
that generate the external field. The magnets are aligned parallel to each other at a distance
$2R= 120$~mm and create an external
magnetic field of magnitude $\mu_0 H_\text{ext}= 45$~mT that penetrates the two-dimensional
pattern, the steel sphere, the wax/magnetite spheres, and the ferrofluid droplets. Dipolar
interactions between two wax/magnetite spheres or between ferrofluid nano particles
are weak as compared to the interaction with the pattern and the external field. 

The gradient of the magnitude of the external field
at the position of the transported objects 
${\boldmath \nabla H_\text{ext}}\approx M_\text{ext}t_\text{ext}d_\text{ext}^2/R^4$
is at least two orders of magnitude smaller than that of 
the magnetic field of the pattern ${\boldmath \nabla H_{p}} \approx (M_l+M_s)/a$.
Hence, the field created by the external magnets is effectively uniform.
The two external magnets can be oriented to produce an arbitrary direction of the external
magnetic field with respect to the pattern. A laser pointing along ${\mathbf H}_\text{ext}$
is mounted on the goniometer, see Fig.~\ref{fig1}e, to create a stereographic projection of the instantaneous external
magnetic field direction on a recording plane.

\section{Topologically nontrivial transport loops}

\begin{figure*}
	\includegraphics[width=0.95\textwidth]{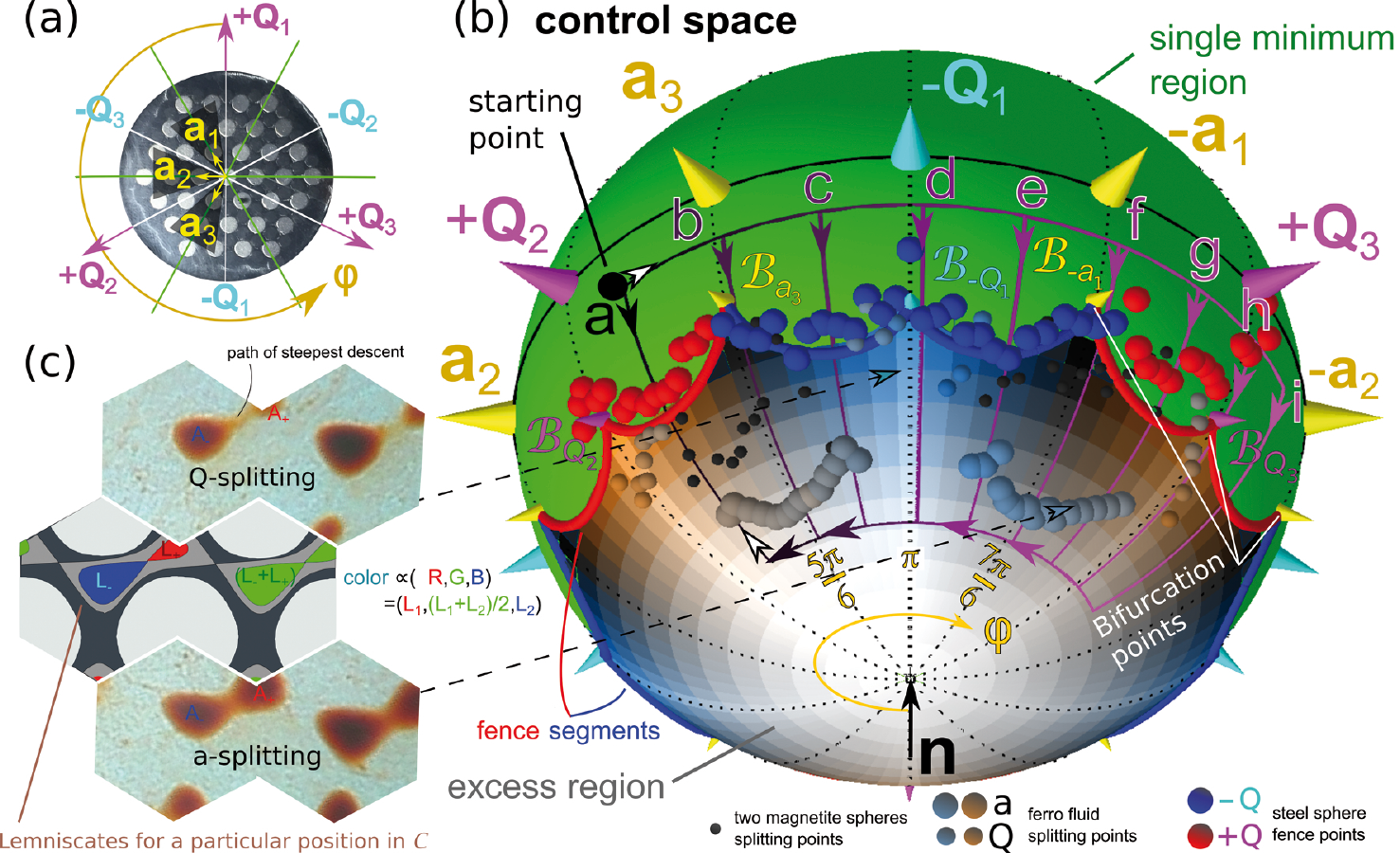}
	\caption{
		\textbf{(a)} Direction of the primitive unit vectors $\mathbf a_1$, $\mathbf a_2$,
		and $\mathbf a_3$ of the direct lattice and direction of the primitive
		unit vectors $\mathbf Q_1$, $\mathbf Q_2$, and $\mathbf Q_3$ of the
		reciprocal lattice. \textbf{(b)} Control space of the hexagonal lattice. Theoretical
		fences between the region of one unique minimum (green) and the excess region
		for paramagnetic objects are shown as red ($+\mathbf Q$ segments) and blue
		($-\mathbf Q$ segments) lines. Experimental fence data from the single steel
		spheres are shown as red and blue spheres. The experiments are performed with
		modulation loops (a-i) that start at the big black circle (starting point)
		and enter the excess region in the south of ${\cal C}$ along a longitude between the
		$\mathbf Q_2$ and the $-\mathbf a_2$ longitudes through
		either a red or blue fence segment. The loops exit this region and return to 
		the starting point through the red fence along a longitude between $\mathbf a_3$
		and $\mathbf Q_2$. We also used the time reversed loop -i  of the loop i.
		We measure the transport of paramagnetic
		objects on the pattern as a function of the entry longitude that we continuously
		vary as a function of the azimuthal angle $\phi_{\text{entry}}$. The experimentally color
		coded spheres 
		show the measured splitting location of ferrofluid droplets and
		of wax/magnetite doublets, see legend. The color decodes the different sizes of the split
		objects according to equation \eqref{EQcolor}. The same color coding with the droplet areas replaced by the sub areas of the theoretical lemniscates is used for the background in the excess region. The unit vector $\mathbf n$ is normal to the pattern.
		\textbf{(c)} Two unit cells with experimental ferrofluid droplets at a $\mathbf Q$-splitting line for $\phi_{\text{entry}}\approx\pi$ (top).
		Two unit cells of the pattern with theoretical lemniscates (equipotential lines through saddle points)
		computed from the magnetic potential for an external field in the excess region (middle).
		Two unit cells with experimental
		ferrofluid droplets at a $\mathbf a$-splitting line  on the loop g for
		$\phi_{\text{entry}}\approx 7.5 \pi/6$ (bottom).
		The dashed arrows point at the corresponding orientations of the external field in ${\cal C}$.} 
	\label{fig2}
\end{figure*}

The parametric dependence of the potential acting on a paramagnetic object (equ. \eqref{EQpotential}),
was studied in detail in Ref.~\cite{Loehr}. The potential has hexagonal symmetry and the number of minima per unit cell of the potential can be one or two, depending on the
orientation of the uniform external field. The set of possible orientations of the external field
forms a sphere that we call the control space $\cal C$ (see Fig.~\ref{fig2}b). 
Two minima exist in the excess region of $\cal C$ (see Fig.~\ref{fig2}b) in which
the orientation of the external field is roughly antiparallel to the magnetization $\mathbf{M}_l$ 
of the silver magnets of Fig~\ref{fig2}a. Only one minimum of the potential exists for orientations
of the external field roughly parallel to the magnetization of silver magnets (see green region of control space in Fig.~\ref{fig2}b).
The boundary in $\cal C$ between the excess region and the region of one single minimum
is a closed curve in $\cal C$ that we call the fence $\cal F$. The fence consists of twelve
segments (red and blue in Fig.~\ref{fig2}b) meeting in twelve bifurcation points.
These bifurcation points (${\cal B}_{\pm a_i}$, and ${\cal B}_{\pm Q_i}$) are located
in the southern hemisphere of $\cal C$ on longitudes running through the directions
$\pm\mathbf a_i$ and $\pm\mathbf Q_i$ ($i=1,2,3$)
of the primitive unit vectors of the direct and reciprocal lattice,
respectively (see Figs.~\ref{fig2}a and \ref{fig2}b). The fence segments
are of two types $+\mathbf{Q}$-segments (red segments in Fig.~\ref{fig2}b) and 
$-\mathbf{Q}$-segments (blue segments in Fig.~\ref{fig2}b).

We reorient the external magnets by moving along a closed reorientation
loop that starts and ends at the same orientation. (See the black point
in Fig.~\ref{fig2}b marked as starting point between the $\mathbf Q_2$ and the
$\mathbf a_3$ longitude in the southern unique minimum region.) As a result
of the reorientation loop of the external field, the steel sphere,
the wax/magnetite spheres, and the ferrofluid droplets move above the magnetic
lattice.
A topologically trivial motion of the steel sphere is that
where the sphere responds to a closed reorientation loop with a
closed loop on the lattice. Not every closed reorientation loop causes such a
trivial response of the steel sphere. There are topologically nontrivial
trajectories, where the steel sphere ends at a position differing
from the initial position by one unit vector of the magnetic lattice. Nontrivial
closed reorientation loops in control space are those loops that have loop segments
in both, the excess region and the region of the unique minimum~\cite{Loehr,Rossi}.
Here, the reorientation loop enters the excess region with a longitude $\phi_{\text{entry}}$ between
the $\mathbf Q_2$ and the $-\mathbf a_2$ directions and exits the excess region
between the $\mathbf Q_2$ and the $\mathbf a_3$ longitudes at $\phi_{\text{exit}}=4.4\pi/6$.
Schematic reorientation loops in
control space are depicted in Fig.~\ref{fig2}b.

\subsection{Single steel sphere transport}
We have reported in Refs.~\cite{Loehr,Rossi,Loehr2} how the transport of a single paramagnetic particle
changes as we move the entry longitude $4.4\pi/6 <\phi_{\text{entry}}<8.5\pi/6$
of the reorientation closed loop. Here we briefly
repeat the findings which are important for this work. The steel sphere adiabatically
returns to its initial position (performs a closed loop above the lattice) if the reorientation
loop enters and exits the excess region via the same fence segment.
As an example of such a loop we have drawn the loop a
in Fig.~\ref{fig2}b for which $\phi_{\text{entry}}=\phi_{\text{exit}}=4.4\pi/6$. The sphere also
returns to the same position when the reorientation loop encloses only the
bifurcation point ${\cal B}_{\mathbf a_3}$ or the two bifurcation points  ${\cal B}_{\mathbf a_3}$ and ${\cal B}_{-\mathbf Q_1}$ in $\cal C$ such in the case of loops b-e
($5\pi/6 <\phi_{\text{entry}}<7\pi/6$ and $4\pi/6<\phi_{\text{exit}}<5\pi/6$) in Fig.~\ref{fig2}b.
However, as the modulation loop encloses ${\cal B}_{\mathbf a_3}$ the
motion is no longer adiabatic. Instead, an irreversible ratchet jump
occurs as the modulation loop exits the excess region through the fence.
We have measured the position of the fence in control space via these
ratchet jumps. The blue and red spheres in Fig.~\ref{fig2}b are the experimental
data of those measurements. The motion is always adiabatic if the control loop enters
and exits the excess region via the same type of fence segments, either $+\mathbf{Q}$-segments
or $-\mathbf{Q}$-segments~\cite{Loehr,Rossi}. A ratchet occurs if the entry and the exit fence segments
are of different type.
The motion of the steel sphere becomes adiabatic when
enclosing the third bifurcation point ${\cal B}_{-\mathbf a_1}$ (loops f-i
$7\pi/6 <\phi_{\text{entry}}<9\pi/6$ in Fig.~\ref{fig2}b) with a total displacement
of the sphere by one unit vector $-\mathbf a_2$. Since this reorientation loop is adiabatic, the time reversed loop
(e.g. the inverse loop -h with $4\pi/6<\phi_{\text{entry}}<5\pi/6$ and $7\pi/6 <\phi_{\text{exit}}<8\pi/6$)
transports into the opposite, i.e. the $\mathbf a_2$ direction than the direct loop.

\begin{figure*}
	\includegraphics[width=1.0\textwidth]{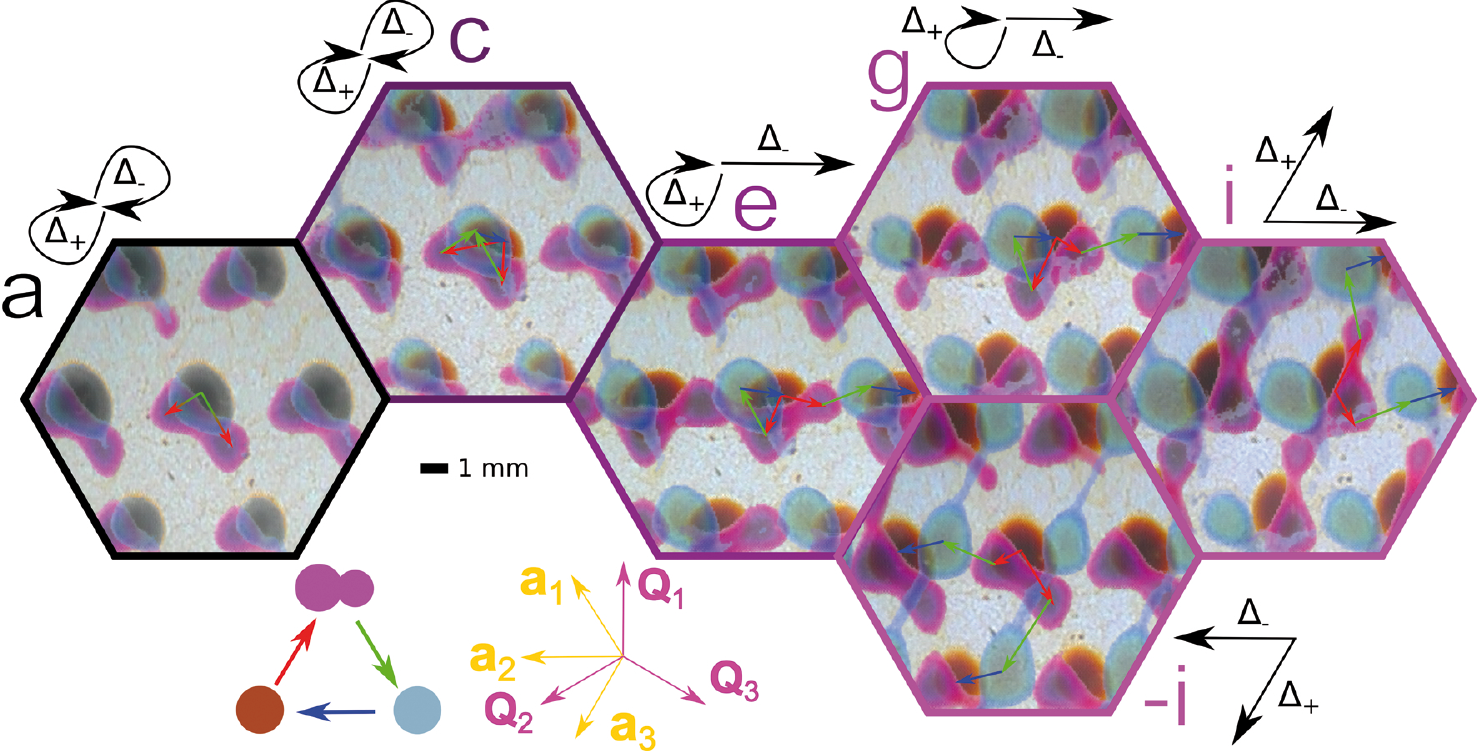}
	\caption{Dynamics of ferrofluid droplets. In each image we overlay an
		image of the droplet before the entry into the excess region (brown),
		at the splitting line (purple) and after recombination at the end
		of the loop (turquoise). The different images correspond to different loops depicted
		in Fig.~\ref{fig2}b with loop a
		$\phi_{\text{entry}}=\phi_{\text{exit}}=4.4\pi/6$, loop c $\phi_{\text{entry}}=5.5 \pi/6$, $\phi_{\text{exit}}=4.4 \pi/6$
		, loop e  $\phi_{\text{entry}}=6.5 \pi/6$, $\phi_{\text{exit}}=4.4 \pi/6$, loop  g $\phi_{\text{entry}}=7.5 \pi/6$,
		$\phi_{\text{exit}}=4.4 \pi/6$ loop i $\phi_{\text{entry}}=8.5 \pi/6$ $\phi_{\text{exit}}=4.4 \pi/6$ and the inverse
		loop -i $\phi_{\text{entry}}=4.4 \pi/6$, $\phi_{\text{exit}}=8.5 \pi/6$. For loops a and c two
		trivial modes coexist. In the loops e and g a transport mode into the $-\mathbf a_2$ direction
		coexists with a trivial mode. In loop i a transport mode in the $-\mathbf a_2$ direction
		coexists with a transport mode into the $-\mathbf a_3$ direction. The control loop
		of -i is the inverse of the loop i with two  transport modes along
		$\mathbf a_2$ and $\mathbf a_3$. The red arrows show the
		transport directions during the splitting of the brown droplet towards the purple droplets.
		The green arrows show the motion of the two purple droplets as they rejoin
		into the turquoise droplet. The blue arrow shows the adiabatic
		motion upon closing the loop in control space by returning to the starting point. The
		displacement after one control loop is the coexistence of the two displacements $\mathbf \Delta_+$
		and $\mathbf \Delta_-$ of the two split droplets. The black arrows are sketches of the
		motion of the droplets.
		The scale bar is $1$ mm.
		A video clip of the motion of the droplets is provided in the supplemental material (adfig3.mp4).}
	\label{fig3}
\end{figure*}

\subsection{Ferrofluid droplet transport}

\begin{figure}
	\includegraphics[width=\columnwidth]{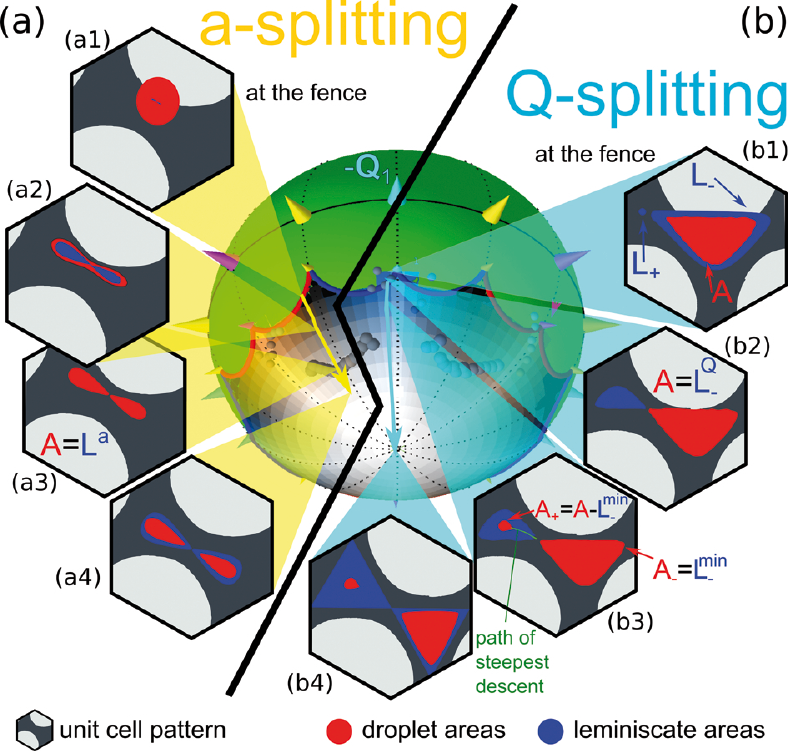}
	\caption{Schematic of the $\mathbf a$-splitting (a) and 
		the $\mathbf Q$-splitting (b) of a ferrofluid droplet. In each case
		a unit cell (hexagon) with the droplets (red) and the lemniscates (blue)
		is represented for four different orientations of the external field along
		a control loop  that enters the excess regions of ${\cal C}$ 
		near a ${\cal B}_a$-bifurcation point (a1)-(a4) and a ${\cal B}_{-Q}$-bifurcation point (b1)-(b4). 
		The loop segments of both loops in control space are indicated in the
		control space in the center of the image. 
		The positions of the large (small) magnets in the unit cell are shown in light (dark) gray.}
	\label{fig4}
\end{figure}

We consider next the motion of ferrofluid droplets since they mimic 
the transport in a system with hundreds of tiny particles per unit cell.
Hence, the transported object can no longer be considered a point particle.
To understand the motion we need to consider the equipotential lines around
the minima of the total magnetic potential that drives the motion.
In Fig.~\ref{fig3} we collect images that show the motion of a ferrofluid
droplet along some of the control modulation loops displayed in Fig.~\ref{fig2}.
The loops enclose from zero up to four bifurcation points. At the starting point
the steel sphere and the ferrofluid droplets reside above the central magnet of
the unit cell. The sphere/droplet moves away from
this location as the external magnetic field enters into the excess region of ${\cal C}$.
Nothing special occurs when the external field crosses the fence, and nothing particular happens
to the single steel sphere as the field moves deeper into the excess region. The
ferrofluid droplet, however, deforms into a dogbone-like shape and eventually
splits in two smaller droplets when the modulation loop crosses a droplet splitting line
in ${\cal C}$. Some of the shapes of such droplets are shown in Fig.~\ref{fig2}c.
Their shape and size agree very well with the shape and size of the lemniscates,
which are simply the equipotential lines of the colloidal potential passing through the
saddle point between both minima, see Fig.~\ref{fig2}c.
The two separated ferrofluid droplets reside in a region above the two different minima
that the potential has in the excess region of ${\cal C}$. The droplets are in general
transported into different directions. When the modulation loop is closed (returns to
the starting point) the two split ferrofluid droplets must return to either the original position
above the same central silver magnet or to an equivalent position in a different unit cell.
The transport over one period is therefore the coexistence of two different types of
transport directions.
The total transport is the sum of the two coexisting displacements weighted
with the two areas of the droplets when they split.

The splitting of a ferrofluid droplet occurs either in an adiabatic way
($\mathbf a$-splitting) or irreversibly ($\mathbf Q$-splitting).
Both types of splitting are schematically represented in Fig.~\ref{fig4}.
The ferrofluid droplet covers a certain area $A$ of action space when
the external field enters the excess region of ${\cal C}$.
A "minor" excess minimum and an excess saddle point are created in the magnetic potential
upon the entry of the external field into the excess region~\cite{Loehr}.
The equi-potential line passing through the excess saddle point is a lemniscate
that first winds around the preexisting "major" minimum, then passes
through the saddle point, and next winds around the minor excess minimum.
Hence, the lemniscate defines a closed curve of area $L=L_++L_-$ where
each of the two sub areas, $L_+$ and $L_-$, surrounds a minima of the potential.
At the fence in ${\cal C}$, the sub lemniscate area of the minor minimum, $L_+=0$ ($L_-=0$)
for a $-\mathbf Q$- fence segment ($+\mathbf Q$-fence segment) vanishes.
At the fence  $L_+=0$ or $L_-=0$ and the area occupied by the ferrofluid 
droplet can be either larger $A>L_++L_-$ or smaller $A<L_-$ ($A<L_+$) than
that of the preexisting major minimum. 

In the case  $A>L_++L_-$ ($\mathbf a$-splitting),
which occurs 
if the loop enters the excess region of ${\cal C}$ close to a ${\cal B}_a$
bifurcation point, the ferrofluid droplet assumes the shape of an equipotential
line containing both minima (see Fig.~\ref{fig2}c bottom and Fig.~\ref{fig4}a).
When the loop enters deeper into the excess region
of $\cal C$, the area of the lemniscate grows, Figs.~\ref{fig4}a1 and~\ref{fig4}a2. At the point where the area of the lemniscate is
the same as the area of the droplet, $L=L^a=A$,  both lemniscate subareas $L^a_+$ and $L^a_-$
are fully filled with ferrofluid, Fig.~\ref{fig4}a3. When the area of the lemniscate grows beyond that of the droplet,
$L>A$, then the droplet splits into two droplets of areas $A_+=L^a_+$ and $A_-=L^a_-$, Fig.~\ref{fig4}a4.
The areas of both droplets do not change until the droplets coalesce again, i.e. no further splitting
occurs. The
splitting is reversible if the control loop is reversed and crosses the splitting line
$L^a=A$ at exactly the same point. In Fig. \ref{fig2}b we have color coded
the splitting ferrofluid droplets with the a normalized RGB-color given by the triplet 
\begin{equation}
\text{(R,G,B)}=\frac8{A_{\text{UC}}}(A_+,(A_++A_-)/2,A_-),\label{EQcolor}
\end{equation}
where $A_{\text{UC}}$ is the area of the unit cell, and the factor eight accounts for the
fact that the maximum subarea of a lemniscate is one eighth of the area of the unit cell (see Fig.~\ref{fig4}b4 ).
We have also color coded the excess region of ${\cal C}$ in Fig.~\ref{fig2}b with the
same criterion but replacing the subareas $A_{\pm}$ with the subareas of the lemniscates $L_{\pm}$.
Hence, the color of the experimental data points at the splitting lines in Fig.~\ref{fig2}b match the color
of control space only if $A_{\pm}=L_{\pm}$, i.e., if the subareas of the theoretical lemniscates and those
of the droplets are equal.
In Fig. \ref{fig2}b the color of the
experimental $\mathbf a$-splitting points are darker than the background indicating
that the experimental droplets split later than predicted by the theory. This is presumably
because adhesive forces of the droplet prevent an early splitting. The color
discrepancy of the data below the ${\cal B}_{-a_1}$ bifurcation points is likely due
to scatter in the magnetization of the $Ni B$ magnets forming the pattern. The finite
size of the pattern and the elevation of the particles above the pattern have
also an effect on the experimental measurements. 

In the case  $A<L_-$ ($-\mathbf Q$-splitting), which occurs if the control loop
enters the excess region close to a ${\cal B}_{-Q}$ bifurcation point,
the ferrofluid droplet assumes the shape of an equi-potential line
surrounding only the preexisting major minimum but not the excess minor minimum.
We show in Fig.~\ref{fig2}c top the picture of a droplet just after a $-\mathbf Q$-splitting (see also
Fig.~\ref{fig4}b).
The largest areas of $L_-$ occur if the external field points at the south pole and
at the ${\cal B}_{-Q}$ bifurcation points of ${\cal C}$. Hence, 
a locally minimal area $L_-^{\text{min}}$ occurs for external fields pointing
along a longitude that connects the south pole and one of the ${\cal B}_{-Q}$ bifurcation points.
The subarea  $L_-$ of the preexisting minimum of the lemniscate shrinks as the control
loop moves from the fence towards the south pole, Fig.~\ref{fig4}b1.
At some point, the subarea of the lemniscate $L_-$
equals that of the droplet $L_-^Q=A$ (provided that $L_-^{\text{min}}<A<L_-$), Fig~\ref{fig4}b2.
There, the fluid completely fills the subarea $L_-^Q$
while the other subarea $L_+$ is completely empty.
When the major area of the lemniscate shrinks below the
droplet area, i.e. $L_-<A$, then the droplets split into two droplets of
areas $A_-=L_-$ and $A_+=A-L_-$, Fig.~\ref{fig4}b3.
The fluid in $L_-$ is
expelled from the droplet through the saddle point and flows down the
path of steepest descend into the basin of the minor excess minimum.
The areas of the droplet change until the decrease of $L_-$ stops, Fig.~\ref{fig4}b4.
The splitting process is irreversible and both droplets cannot be rejoined in a reversible
way since the fluid in the excess minimum cannot flow up the path of
steepest decent back into the preexisting minimum. We have placed
$\pm\mathbf Q$-splitting experimental points at the location where
the splitting starts. The points are colored according to Eq.~\ref{EQcolor}
with $A_{\pm}$ being here the subareas when interchange of fluid between $A_+$ and $A_-$ stops.
The agreement with the theoretical prediction given by the areas of the lemniscate (colored background)
is excellent.

The splitting lines $L^a=A$, $L^Q_-=A$ and $L^Q_+=A$ are segments of
a closed curve that are joined at the fence of $\cal C$. For $A>L_-^{\text{min}}$
any closed curve of a modulation loop in $\cal C$ that penetrates the excess
region  deep enough must also pass the splitting curve. A nontrivial modulation
therefore causes nontrivial transport that is the coexistence of two
topological displacements weighted with the two split areas of the
droplet. The splitting areas $A_-$ and $A_+$ continuously change along
the splitting lines. A video clip (adfig3.mp4) showing more details of the ferrofluid  transport is provided in the supplementary information.

\subsection{Doublet transport}
We have also studied experimentally the transport of two particles per unit
cell. The area $A$ enclosed by the two wax/magnetite spheres is smaller than the local minimum
area, i.e. $A<L_-^{\text{min}}$. Hence, in contrast to the ferrofluid droplet the $\mathbf Q$-splitting cannot
occur for these doublets. Both, the ferrofluid droplet
and the doublet exhibit $\mathbf a$-splitting.
A measurement of the $\mathbf a$-splitting line $L^a=A$ is shown for the wax/magnetite
doublets in Fig. \ref{fig2}b. The doublets 
are transported together within the major minimum if the control loop enters the
excess region in the vicinity of a ${\cal B}_{\pm Q}$ bifurcation point. If the loop
enters the excess region in the vicinity of a ${\cal B}_a$ bifurcation point, then an $\mathbf
a$-splitting occurs and separates both spheres. One sphere is transported within the
major minimum and the other one within the minor minimum. Two transport directions
coexist. 

Hence, depending on $\phi_{\text{entry}}$ there are two different transport modes for the
doublets: (i) no splitting and (ii) $\mathbf a$-splitting.
The transition from one transport mode towards the other 
happens at doublet bifurcation points ${\cal B}_{\text{doublet}}$ that are the
intersections of the fence with the $\mathbf a$-splitting line $L^a=A$ for doublets. 
Since the spheres have the same size, the areas of the $\mathbf a$-splitting transport
are equal $A_-=A_+=A/2$. Hence, the experimental data
points for doublet $\mathbf a$-splitting have all the same color, see Eq.~\eqref{EQcolor}
and Fig.~\ref{fig2}b. The color is darker than the theoretical background color of the lemniscates.
Hence, the splitting happens later in the experiment than predicted by the theory, presumably due to
dipolar attraction between the two spheres as well as due to friction with 
the bottom surface. Like the transport of a single sphere, the transport of
doublets is discrete and therefore topological. 


\begin{figure*}
	\includegraphics[width=0.9\textwidth]{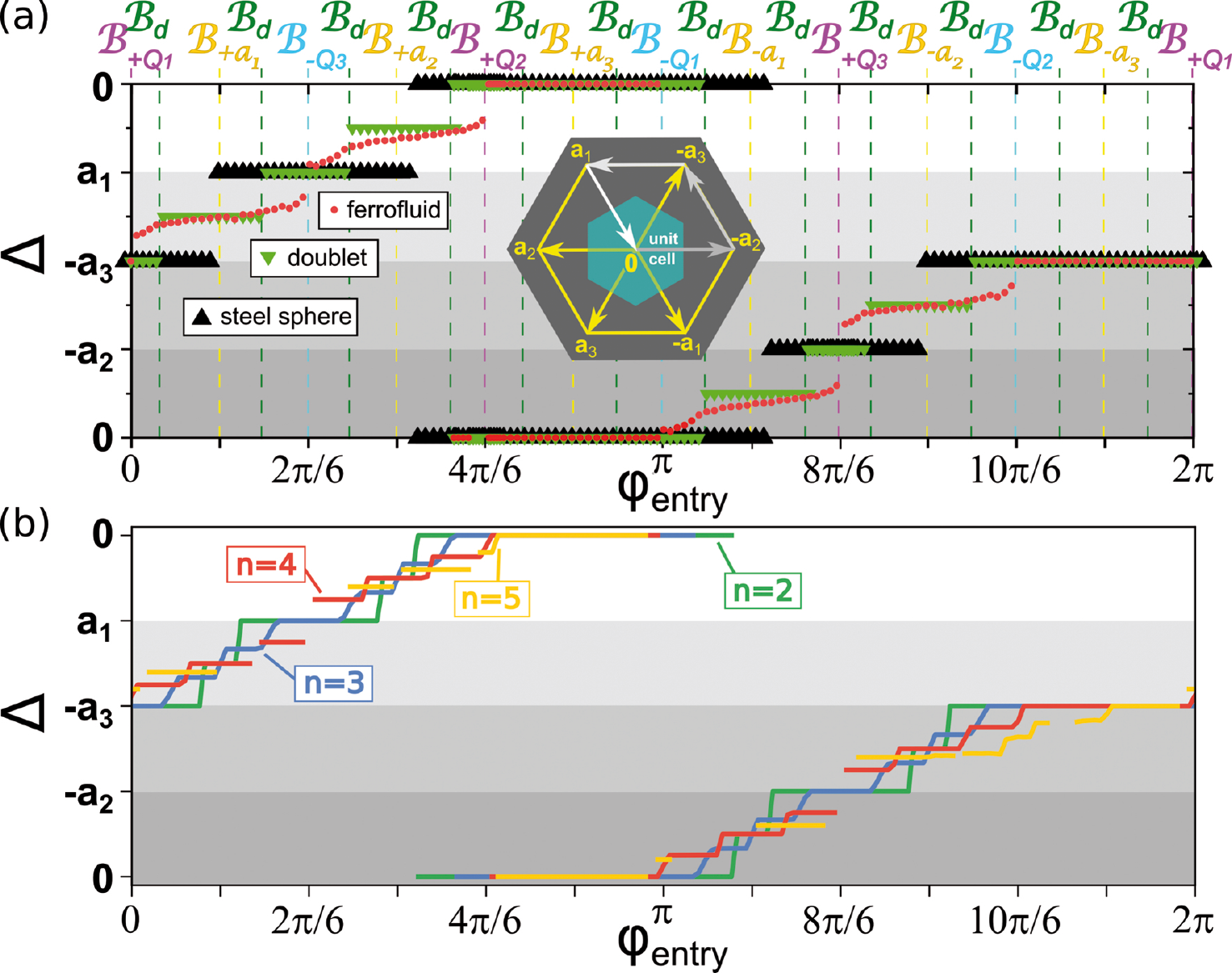}
	\caption{(a) Experimental measurements of the net displacement $\mathbf \Delta$ of the steel sphere,
		the wax/magnetite doublet, and the ferrofluid droplets as a function of $\phi_{\text{entry}}$ for a family
		of loops with  $\phi_{\text{exit}}=4.4\pi/6$. The displacement changes along the four gray arrows shown
		in the the center of the image. The four gray shaded regions correspond to the arrows of the same color
		in the inset. The displacement is a discrete function of $\phi_{\text{entry}}$ for the steel speher and the wax/magnetite
		spheres but a continuous function for the ferrofluid droplets. The jumps in the displacement occur when the
		loop crosses the bifurcation points in control space (the position of which is indicated with vertical dashed lines).
		A video of the motion of a steel sphere, a wax magnetite doublet, as well as ferrofluid droplets subject to two
		different control loops is shown in the supplemental file adfig5.mp4.
		(b) Net displacement of a collection of $n=2,3,4$, and $5$ particles, as indicated, for the same family of control
		loops as in panel (a) according to computer simulations.}    
	\label{fig5}
\end{figure*}

{\bf Net displacement} We analyze next the net displacement after the completion of one entire control loop for all three types of objects: single spheres, doublets, and ferrofluid droplets.
For all objects, we define the vector of the net displacement $\mathbf \Delta$ as the area averaged sum of the two possible displacements. That is
\begin{equation}
\mathbf \Delta=\frac{\mathbf \Delta_+ A_+ + \mathbf \Delta_- A_-}{ A_+ + A_-}, 
\end{equation}
where $\mathbf{\Delta}_\pm$ are the net displacement vectors of the two minima in one control loop.
Hence, $\mathbf{\Delta}_\pm$ are always lattice vectors and $\mathbf{\Delta}$ moves along a straight line between the two lattice vectors.

In Fig. \ref{fig5}a we plot the net displacement of all magnetic objects
as a function of $\phi_{\text{entry}}$ for a family of loops with
$\phi_{\text{exit}}=4.4\pi/6$ (loops similar to those in Fig.~\ref{fig2}b).
A video of the motion is presented in the supplemental material (adfig5.mp4).
The net displacement is zero when the loop encloses no bifurcation point.
The displacement moves along the straight lines connecting the sequence of lattice vectors
$\mathbf 0$,$-\mathbf a_2$, $-\mathbf a_3$, $\mathbf a_1$, and $\mathbf 0$
(see the gray arrows in the center of Fig. \ref{fig5}a)
The areas $A_+$ and $A_-$ are continuous functions of $\phi_{entry}$ for the ferrofluid droplet.
In contrast, the transported areas $A_\pm$ for one steel sphere (a wax/magnetite doublet) can be only
integer multiples $n=0,1$ ($n=0,1,2$) of the area of one sphere. 
Contrary to the ferrofluid transport, the sphere and doublet net displacement change discretely
with $\phi_{\text{entry}}$. The number of discrete steps for the doublets are twice that of a single steel sphere.   
Hence the transport of one or two spheres is topological while the transport of the ferrofluid is geometrical.

\subsection{Multi particle transport}
What is the minimum number of particles required for having geometrical transport?
To address this fundamental question, we have simulated the transport of multiple spheres using overdamped Langevin dynamics
(note that inertial effects are negligible).
Each unit cell is filled with exactly $n$ particles. Each particle is subject to the magnetic potential
\begin{equation}
U(\mathbf{x}_\mathcal{A}, \mathbf{H}_\text{ext}(t)) \propto -\mathbf{H}_\text{ext}(t) \cdot \mathbf{H}_p(\mathbf{x}_\mathcal{A}),
\end{equation}
where \(\mathbf{H}_\text{ext}(t)\) is the external field at time \(t\) and
\(\mathbf{H}_p(\mathbf{x}_\mathcal{A})\) is the magnetic field created by
the pattern at position of the particle \(\mathbf{x}_\mathcal{A}\) in action space \(\mathcal{A}\),
see Refs.~\cite{delasHeras,Loehr2}.

The particles interact via the purely repulsive Weeks-Chandler-Andersen potential
\begin{equation}
\phi(r) =
\begin{cases}
4 \epsilon \left(\left(\frac{\sigma}{r}\right)^{12}-\left(\frac{\sigma}{r}\right)^6 + \frac{1}{4}\right)&\quad r \leq 2^{1/6}\sigma\\
0&\quad r > 2^{1/6}\sigma
\end{cases},
\end{equation}
where $r$ is the distance between the particles, $\epsilon$ fix our unit and energy, and $\sigma$ is the effective particle length that 
we fix to $\sigma/a = 0.2$ which is the same as in the experiments (transport of single spheres).
We integrate the equations of motion with a time step  $dt/T=10^{-5}$, with $T$ the period of a modulation loop.
Fig. \ref{fig5}b shows the net displacement of $n=2,3,4,5$ spheres. The number 
of plateaus in the displacement of $n$ particles per unit cell is $n$ times the number of plateaus of a single sphere.
This is true provided that each unit cell is filled with precisely $n$ particles since
the results depend on the initial distribution of particles among the different unit cells. Our simulation
results suggest therefore that the change from topological towards geometrical transport is
determined by the precision of the experimental setup and measurements.

Note that the $\phi_{\text{entry}}$ of the doublet bifurcation points (at which the net displacement jumps) differs
in experiments, Fig.~\ref{fig5}a, and simulations, Fig.~\ref{fig5}b. In the experiment, the spheres do not only interact via excluded volume interactions,
but are also subject to long range dipolar interactions. We tried to minimize the effect of dipolar interactions by using the wax/magnetite
composite spheres. The transport properties depend also on the size of the particles, and hence a depth understanding
of how the transport changes from topological to geometrical requires further studies.

\section{Conclusions} 
We have studied experimentally and with computer simulations the transport of paramagnetic particles on top of a magnetic lattice and driven by
a uniform and time-dependent external magnetic field. The external field performs periodic closed loops.
We have shown that increasing the number of particles
within the unit cell of the lattice changes the transport from topological towards geometrical. 

The transport as a function of a parameter that continuously characterizes a family of control loops
is discrete for low particle densities and continuous for a macroscopic number of particles per unit cell (ferrofluid droplet). 
The possibility to split or disjoin soft matter assemblies increases the number of transport modes and bifurcations,
effectively changing the transport from topological towards geometrical.

\section{Acknowledgments}
Scientific discussion with Michel L\"onne is highly appreciated.

\end{document}